\documentclass[fleqn,usenatbib,useAMS]{mnras}

\usepackage{graphicx}	
\usepackage{amsmath}	
\usepackage{amssymb}	
\usepackage{multicol}        
\usepackage{bm}		
\usepackage{pdflscape}	

\usepackage[T1]{fontenc}
\usepackage{ae,aecompl}

\usepackage{newtxtext,newtxmath}

\title[Infrared variability of dusty white dwarfs]{Most white dwarfs with detectable dust discs show infrared variability}

\author
[A. Swan et al.]
{Andrew Swan$^{1}$\thanks{E-mail: \href{mailto:a.swan.17@ucl.ac.uk}{a.swan.17@ucl.ac.uk}},
Jay Farihi$^{1}$, and
Thomas G. Wilson$^{2}$
\\
$^{1}$Department of Physics~\& Astronomy, University College London, Gower Street, London~WC1E~6BT, UK\\
$^{2}$Isaac Newton Group of Telescopes, E-38700 Santa Cruz de~La~Palma, Spain\\
}

\date{Accepted XXX. Received YYY; in original form ZZZ}

\pubyear{2019}

\begin{document}
\label{firstpage}
\pagerange{\pageref{firstpage}--\pageref{lastpage}}
\maketitle

\begin{abstract}
Archival data from the \textit{WISE} satellite reveals infrared flux variations of tens of per cent around numerous dusty white dwarfs.  Data spanning more than seven years reveal more than half of known systems are varying in the {3.4\,\micron} band, while the {4.6\,\micron} data are challenging to interpret due to lower signal-to-noise. The sparsely-sampled data limit interpretation, but the heterogeneous light curves suggest each source may be idiosyncratic, where there may be competing processes operating on different time-scales. Collisions are likely driving the observed decays in flux, and this finding suggests that dust production is operating more often than indicated by previous observations. The observed variation is at odds with the canonical flat disc model in isolation, and underscores the need for infrared monitoring of these evolved planetary systems to inform the next generation of theoretical models.
\end{abstract}




\section{Introduction}
\label{sectionIntroduction}

White dwarfs often display photospheric pollution by metals attributed to the accretion of material from their remnant planetary systems (see e.g. the review by \citealt{Jura2014}). Over 30~years ago an infrared excess was discovered at one such star, WD\,2326+049\footnote{Stars are referred to throughout by their WD~designation (their abbreviated B1950.0 coordinates). The remainder of the Letter uses the numerical designations alone, with other names in common use noted where appropriate.}~($=\text{G29-38}$), that was shown to be consistent with a dust disc rather than a cool companion \citep{Zuckerman1987, Graham1990}. Launched in 2003, the \textit{Spitzer Space Telescope} enabled the discovery of over 25 new systems, confirming the link between these orbiting debris discs and photospheric pollution \citep{Farihi2016}.

For 15~years the standard model for the infrared emission has been that of a geometrically thin, optically thick disc akin to the rings of Saturn where the debris orbits within the stellar Roche limit \citep{Jura2003}. Delivery of solids towards the white dwarf has been modelled successfully by Poynting-Robertson (PR) drag \citep{Rafikov2011}, with disc lifetimes exceeding $10^{5}$\,yr predicted by theory and supported by observations \citep{Bochkarev2011, Girven2012}. In this model, accretion is fed only from the narrow, optically thin annulus at the inner edge of the disc, and therefore rapid evolution is not expected, since the disc lifetime is directly proportional to its mass. If on the other hand the discs are optically thin, then the disc lifetime is equal to the PR~drag timescale, which depends instead on the stellar luminosity. Micron-sized dust within the Roche limit of a white dwarf is removed within around 10\,yr, where the time-scale increases linearly with particle size \citep{Hansen2006}.

In the context of the optically thick disc model it was therefore surprising when a drop of 35\,per~cent was seen in the infrared flux from the dusty white dwarf 0956$-017$ ($=\text{SDSS\,J095904.69}-020047.6$) within 300\,days \citep{Xu2014}. All dusty white dwarf systems with gaseous emission lines exhibit optical variability in the circumstellar lines \citep{Gansicke2008}, and appear to be varying in a manner consistent with eccentric rings precessing under general relativity \citep{Manser2016a, Cauley2018, Dennihy2018, Miranda2018}. While this type of optical variability is consistent with geometric changes, short-term changes in dust production are manifest through the transiting events towards 1145+017, which evolve on time-scales of days \citep{Vanderburg2015, Gansicke2016}. To date, no unambiguous changes in the strength of the photospheric metal absorption features have been observed in a polluted white dwarf. However, viscous spreading of gas likely drives accretion onto the stellar surface, and gaseous discs are expected to evolve on time-scales no shorter than decades \citep{Metzger2012,Farihi2018GD56}.

Variation in the infrared regime allows insight into the processes producing and removing dust around the star, where flux changes result directly from an increase or decrease in the emitting surface area. Only one example -- 0956$-017$ -- was known until the recent report of variations at three additional systems \citep{Farihi2018GD56, Xu2018}, where 0408$-041$ ($=\text{GD\,56}$) is the first case of both brightening and dimming events, that took place over several years. This study reports on light curves from the \textit{Wide-field Infrared Survey Explorer} (\textit{WISE}) for 35~stars that reveal most dusty white dwarfs vary at infrared wavelengths. The sample and data retrieval are discussed in Section~\ref{sectionObservations}, the results of the analysis are presented in Section~\ref{sectionResults}, and their interpretation and implications are discussed in Section~\ref{sectionDiscussion}.

\section{Sample selection and source data}
\label{sectionObservations}

Since early 2010, \textit{WISE} has been performing a space-based all-sky infrared imaging survey, the deepest of its kind \citep{Wright2010}. The initial mission lasted 12~months and in late 2013 the spacecraft was revived for the NEOWISE reactivation mission, which remains ongoing \citep{Mainzer2011, Mainzer2014}. \textit{WISE} scans great circles on the sky, making approximately 15~revolutions per day while precessing 360\,\degr\,yr$^{-1}$ to cover the whole sky twice per year. Sources are measured multiple times during each biannual pass (referred to throughout the Letter as an epoch), with the exact number depending on their position but typically around 12 per epoch. Photometry produced by the data reduction pipeline is made publicly available periodically. The most recent data release in 2018~April provides coverage through 2017~December.

White dwarfs with infrared excess consistent with dust discs published in the literature provide up to 45 candidates for inclusion in the sample. The AllWISE Multiepoch Photometry Table and the NEOWISE-R Single Exposure Source Table\footnote{Available at {\href{https://irsa.ipac.caltech.edu}{irsa.ipac.caltech.edu}}} were queried using a 5\,arcsec search radius, accounting for proper motion. The spacecraft detectors operated initially in four bands, but only the {3.4\,\micron}~(\textit{W1}) and {4.6\,\micron}~(\textit{W2}) bands remain operable since depletion of the cryogen. Measurements in these two shorter-wavelength bands, where white dwarf discs are bright and detectable, were retrieved; the \textit{W3} and \textit{W4} bands were not used. 

All measurements marked as upper limits were discarded, as were spurious outliers in position or flux. The AllWISE Reject Table was consulted to exclude the false duplicate measurements that can arise for example when a source features in two overlapping frames. The documentation highlights numerous issues that may affect the data, such as lower-quality frames, scattered moonlight, and image artefacts. Any measurements thus flagged were not included in the analyses.

Some of the target stars have proper motions above 0.2\,arcsec\,yr$^{-1}$, but this is not a concern as the AllWISE photometry solution takes account of linear motion between frames, while the NEOWISE pipeline measures sources directly from individual images. Contamination by nearby sources is a concern because the \textit{WISE} point-spread function (PSF) has a full-width-at-half-maximum (FWHM) of 6.1\,arcsec in the \textit{W1}~band. \textit{Spitzer} IRAC stellar images in comparable wavebands have 2.0\,arcsec FWHMs and exist for all but one of the targets; these images were retrieved from the archive and examined for potential sources of photometric contamination. Targets with comparably bright sources within several arcseconds were rejected if their AllWISE catalogue positions were discrepant beyond the $3\upsigma$~level with their expected locations based on \textit{Gaia} astrometry. Additionally, the flux measurements (but not the errors) of the the retained targets were diluted to account for potential contribution from neighbouring sources (conservatively estimated by reference to the PSF curve of growth provided in the \textit{WISE} documentation). Only around 1~per~cent of field sources are seen to vary by \textit{Spitzer}, so variation by potential contaminants is not a concern \citep{Kozowski2010,Polimera2018}. A total of 35 dusty stars remain in the sample for study.

Most stars have data from ten epochs available in total, two or three of which are from the cryogenic mission. With just a few exceptions, the data cover baselines in the range 7.0--7.5\,yr. The median signal-to-noise ratio (S/N) of individual measurements within each epoch is 9.3 in the \textit{W1}~band and 5.5 in the \textit{W2} band, and therefore to increase sensitivity to flux changes the weighted means and errors for each epoch are analysed.

In order to ensure that any flux variations in \textit{WISE} data are real, a comparison sample of white dwarfs was analysed in an identical manner. Hydrogen-atmosphere white dwarfs with $T_{\text{eff}}$ in the range 9000--25\,000\,K and masses near 0.6\,$M_{\sun}$ were selected for this purpose \citep{Holberg2008, Gianninas2011}. After rejecting known variable stars, those with problematic \textit{WISE} data, or \textit{Spitzer} images indicating possible sources of flux contamination, 37 comparison stars remain.

\section{Results}
\label{sectionResults}
\begin{figure}
 \includegraphics[width=\columnwidth]{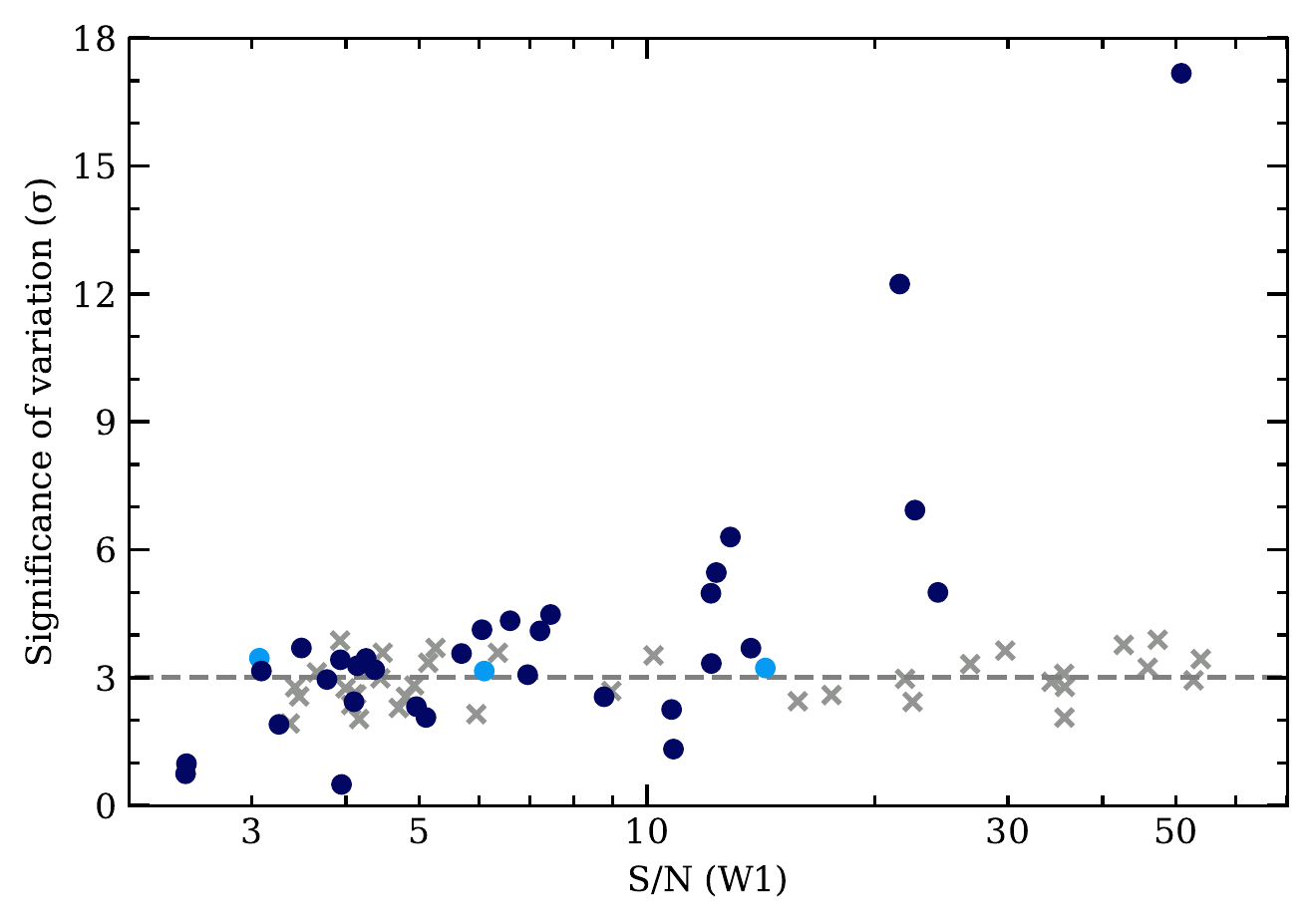}
 \caption{Variation versus signal-to-noise in the \textit{WISE} \textit{W1} band for target and comparison samples. The y-axis shows the level of the most significant detected variation, and the x-axis plots S/N logarithmically for clarity. The horizontal dashed line indicates the $3\upsigma$ level, above which a variation is deemed significant. Dark blue filled circles are science targets, with lighter blue symbols explained in the text, while grey crosses are comparison stars.
 }
 \label{figureVariation}
 \end{figure}

To detect -- rather than characterise -- variation at a star simply requires taking the difference between the mean fluxes at two separate epochs, and calculating the propagated error. If the difference is significant at the $3\upsigma$~level then variation is detected. Only the \textit{W1}~band is considered owing to S/N, and consecutive epochs can be combined to give a single mean. For each epoch, this test is applied systematically against all other epochs (single or combined). Often, but not always, the most significant variation is found to be the peak-to-peak variation in the light curve consisting of single epoch, weighted mean fluxes. The most significant variations for the sample are plotted in Fig.~\ref{figureVariation}, where the lighter blue points represent those targets where variation was only detected if combined epochs were used.

Across the whole sample, 24 of the 35 stars (69~per~cent) show a variation above the $3\upsigma$ level. The sensitivity of the \textit{W1} band exhibits a rapid deterioration in the expected S/N beyond $16^{\text{th}}$ magnitude \citep{Wright2010}. Of the stars dimmer than that magnitude, 56~per~cent vary, while 73~per~cent of those brighter vary. Excluding stars with fewer than 100 measurements, to avoid sparsely sampled objects, leaves 76~per~cent varying, suggesting that the result found for the whole sample is conservative. By contrast, the comparison sample does not show any change in detected variation across these subsamples. These results are summarised in Table~\ref{tableResults}.

\begin{table}
    \caption{Variation statistics in \textit{W1} for the dusty and comparison samples: total in (sub)sample, number, and fraction showing variation above $3\upsigma$.}
    \centering
    \label{tableResults}
\begin{tabular}{lcccccc}
\hline
                    & \multicolumn{3}{c}{...........dusty...........} & \multicolumn{3}{c}{...........comp...........} \\
                    & $N_{\text{tot}}$  & $N_{\text{var}}$  & $f_{\text{var}}$  & $N_{\text{tot}}$  & $N_{\text{var}}$  & $f_{\text{var}}$ \\
\hline
All stars              & 35    & 24    & 0.69  & 37    & 15    & 0.41  \\
$n_{\text{obs.}}>100$  & 25    & 19    & 0.76  & 27    & 12    & 0.44  \\
$m\geq16$\,mag         & 9     & 5     & 0.56  & 7     & 3     & 0.43  \\
$m<16$\,mag            & 26    & 19    & 0.73  & 27    & 12    & 0.44  \\
\hline
\end{tabular}
\end{table}

The errors on measurements reported in the \textit{WISE} data are typically lower than the standard deviation of those measurements within each epoch, suggesting that the errors may be under-reported, and thus some false positives are expected when testing for variability. To illustrate this, consider simulated measurements that are drawn from a normal distribution with a fixed mean and a 10~per~cent standard deviation, but are reported with 2~per~cent errors. Ten epochs, each consisting of ten measurements, will typically lead to a 3.5$\upsigma$ variation being detected. This effect appears to be present in the results for the comparison stars. However, the variation in the comparison stars never exceeds 4$\upsigma$, whereas most variation in the target sample exceeds 4$\upsigma$ for sources with $\text{S/N}>5$.

\begin{figure*}
 \includegraphics[width=0.495\linewidth]{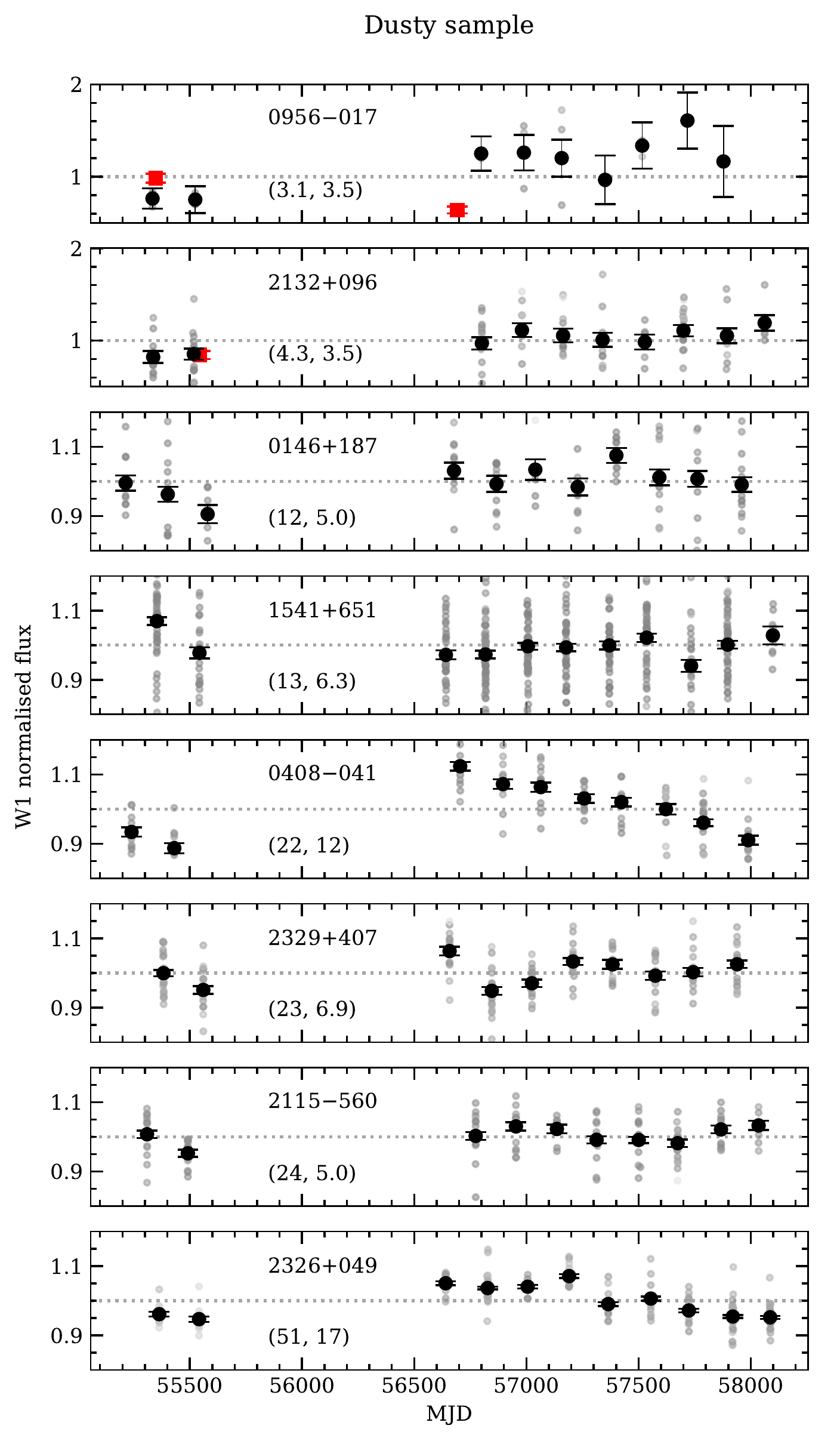}
 \includegraphics[width=0.495\linewidth]{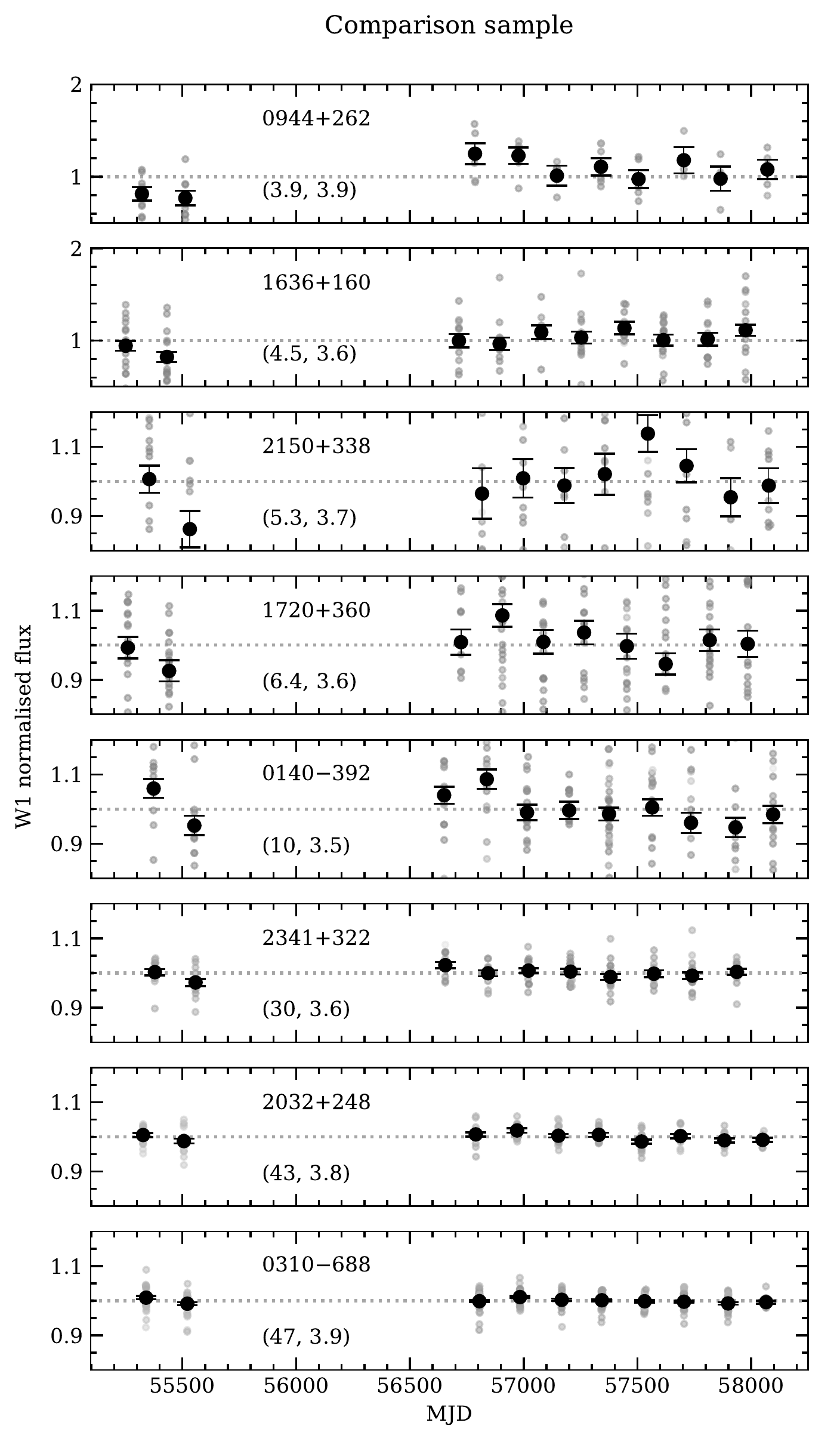}
 \caption{Single epoch light curves for selected stars. The weighted mean and error of each \textit{WISE} epoch are plotted in black. Individual measurements are plotted in grey; some outliers fall beyond the axes limits. Numbers in parentheses beneath the WD designations are their (S/N, $\upsigma$) coordinates in Figure~\ref{figureVariation}. Dusty white dwarfs are shown in the left panel, where red squares (near MJD~55350 and 56700 for 0956$-017$, and MJD~55550 for 2132+096) are the \textit{Spitzer} IRAC measurements \citep{Xu2014, Bergfors2014} discussed in the text. The right panel shows the comparison stars that display the highest apparent variation.}
 \label{figureLightCurves}
\end{figure*}

Fig.~\ref{figureLightCurves} shows the light curves for selected stars and demonstrates a variety of brightening and dimming behaviour. Black points show the weighted means and errors for each biannual epoch, and the grey points show the individual measurements. In contrast to 0408$-041$, most of the stars show more complex behaviour. For example, 1541+651 shows episodes of inter-epoch dimming of 8~per~cent, but also a gradual increase in brightness between. Meanwhile, 2329+407 shows no long-term trend but hints at possible oscillatory variation. Across the sample, between single epochs the significant flux variations in \textit{W1} (as measured against the mean flux) range from 6~per~cent at 0420$-731$ to 72~per~cent at 1145+017, with a mean of 25~per~cent and standard deviation of 16~per~cent. Despite this diversity, a couple of common themes can be identified.

A number of dimming episodes are present, where the gradient of these events is similar across the light curves of several stars, suggesting a single timescale may govern the dust removal process. Following a dimming episode the flux often increases; examples can be seen in Fig.~\ref{figureLightCurves} at 1541+651 (the second dimming event) and 2329+407, where in each case the flux has recovered within two years. This is in contrast to 0408$-041$ where the flux appears to have been in decline since 2014 or earlier. This reveals that dust production is more active, or active more often, than shown by previous observations.

There is no obvious suggestion that any strong variations in colour are occurring. Taking weighted means and errors of the $\text{\textit{W1}}/\text{\textit{W2}}$ ratio for the cryogenic \textit{WISE} and NEOWISE missions respectively finds minimal long-term colour variation: four targets show a colour change above $3\upsigma$, but none vary beyond 4$\upsigma$. However, this test can only rule out flux ratio variations above the 10--20~per~cent level. The stars showing \textit{W1} variations above $5\upsigma$ all have correlated \textit{W1} and \textit{W2} fluxes (particularly 2326+049 which is discussed further below), but otherwise the low S/N of the \textit{W2}~data are such that firm conclusions cannot be drawn from the colour information.

\textbf{0956$-$017} -- This is the first star reported to exhibit long-term variation at infrared wavelengths, showing a flux decrease of around 35~per~cent during 2010, and found to be commensurately low when re-measured in 2014  \citep{Xu2014}. Interestingly, the NEOWISE data indicate the flux has increased since 2014.  While the 41 individual \textit{W1} measurements have insufficient S/N (typically~3) to establish inter-epoch variation, the weighted mean and errors of all cryogenic and warm \textit{WISE} mission data are $67\pm7$\,$\upmu$Jy and $110\pm7$\,$\upmu$Jy respectively. Thus the flux has, on average, nearly doubled during NEOWISE. The background source and potential photometric contaminant reported from higher spatial resolution IRAC images contributes, at most, a stable $9\pm2$\,$\upmu$Jy to the \textit{WISE} fluxes \citep{Xu2014}. Therefore the change in average \textit{W1} fluxes from warm to cryogenic eras is real and at least $43\pm10$\,$\upmu$Jy.

\textbf{2132+096} -- Using \textit{Spitzer} IRAC, this star was found to have a significant excess at {4.5\,\micron} but not at {3.6\,\micron} \citep{Bergfors2014}. The shorter wavelength IRAC and cryogenic \textit{WISE} mean fluxes are similar, at $98\pm5$ and $99\pm5$\,$\upmu$Jy respectively. However, the mean \textit{W1} flux during the NEOWISE mission is $124\pm3$\,$\upmu$Jy, indicating a significant increase in flux, and thus in excess of the stellar photosphere at {3.4\,\micron} during that period.

\textbf{2326+049} -- This is a well-known ZZ~Ceti star \citep{Shulov1974} that pulsates from the ultraviolet \citep{Sandhaus2016}, all the way through to the near-infrared \citep{Patterson1991}, including at {3--8\,\micron} as seen by \textit{Spitzer} IRAC \citep{Reach2009}. The pulsation periods vary from a few to tens of minutes \citep{Kleinman1998} and are almost certainly detected in the \textit{WISE} data. The individual measurements show a strong correlation (Pearson correlation coefficient~0.81) between \textit{W1} and \textit{W2}, indicating that the scatter arises from real variation. The \textit{WISE} integration time of 8.8\,s is more than an order of magnitude shorter than the known pulsation periods and thus the stellar flux remains approximately constant during each measurement. Similar scatter attributed to pulsations in \textit{K}-band and \textit{Spitzer} IRAC measurements of this star have recently been reported \citep{Xu2018}. In principle, the pulsations could be modelled and removed in an effort to search for real variations in the dust disc, but in practice this is not possible due to the plurality of possible simultaneous modes \citep{Kleinman1998}. Thus, in order to potentially disentangle the stellar variation, the response of the dust disc to the thermal pulses from the star, and any separate long-term dust variability, simultaneous observations across a wide range of wavelengths are (minimally) required.

\section{Discussion and conclusions}
\label{sectionDiscussion}

Significant infrared flux variations are seen in 69~per~cent of the known dusty white dwarf sample. The divergence between the target and comparison samples towards higher S/N in Fig.~\ref{figureVariation} is compelling evidence that most dusty white dwarfs are varying at infrared wavelengths, and therefore that the canonical model disc that quiescently feeds metals to white dwarfs is not the complete picture.

One important result that was not previously apparent is that dust production at most stars is an ongoing, stochastic process: while dips appear regularly in the light curves, the flux often recovers afterwards. The challenge now is to identify the process or processes that are creating or recycling dust. A number of competing processes may operate on material in the vicinity of these discs, for example, PR~drag, collisions, tidal disruption, and sublimation, each with a characteristic time-scale. These have been discussed in detail in the context of 0408$-041$ \citep{Farihi2018GD56} but those arguments are universally applicable.

Changes on time-scales of a few years are observed in the infrared and these appear superficially similar to the in-spiral time for micron-sized dust grains under PR~drag \citep{Hansen2006}. However, for example, if acting on a cloud of optically thin dust liberated by an impact, the lack of evidence for temperature changes may argue against it being an important contributor to the infrared variability. Moreover, it has been demonstrated that the collisional time-scale for optically thin dust will always be at least an order of magnitude shorter than the PR~drag time-scale \citep{Farihi2008, Kenyon2017collisions}, meaning that collisional destruction will occur before any significant radial migration. \textit{WISE} provides insufficient constraints on the variation (or lack thereof) in the infrared colour, and thus better data are needed to rule out dust migration as related to the observed flux variability. Changes between the cryogenic \textit{WISE} and NEOWISE colours might be expected to occur on PR~drag time-scales, however, the sensitivity limit for the weighted mean \textit{W1}/\textit{W2} flux ratios found here is typically only 10--20\,per~cent. Such a colour change equates to a commensurate change in temperature around 1000\,K that in turn would translate to a 20--50\,per~cent change in radial distance from the star for optically thin dust. Thus, in the worst case, dust crossing up to half the width of a canonical disc cannot be confidently ruled out.

Simulations show that a collisional cascade of small bodies ($<1$\,km) in an annulus around a white dwarf can effectively destroy solids that are efficient infrared emitters, producing a detectable excess for only a few years \citep{Kenyon2017collisions}. If collisions dominate disc evolution in this way, the required rate of replenishment ($\gtrsim10^{12}$\,g\,s$^{-1}$) is at odds with that typically inferred from observed metal abundances and atmospheric modelling ($\lesssim10^{9}$\,g\,s$^{-1}$; \citealt{Farihi2018GD56}). Stochastic disruption of small bodies may play a role in resolving this tension: transient production of optically thin material may cause time-varying emission that supplements steady emission from a canonical disc.

Because an optically thick disc is effective at damping collisions \citep{Metzger2012}, any material co-orbital with such a disc would be rapidly (re-)absorbed. Therefore any transient dust production that leads to an increase in infrared flux must be separated radially from an optically thick disc. Bodies that are potential sites of dust production could be supplied from further out, forming a ring of debris after disruption that then circularises and gradually shrinks towards a canonical disc \citep{Debes2011,Bonsor2011, Veras2014b}. Alternatively, such bodies could form in situ, perhaps near to but outside the Roche limit, as has been suggested for the rings of Saturn \citep{Charnoz2010, vanLieshout2018}. With sufficient data it would be possible, in principle, to fit analytic models of collisional cascades (e.g.~\citealt{Wyatt2007, Kenyon2017model}) to the decay episodes.

The lack of obvious periodicities in the light curves suggest that interactions with material on wider orbits do not dominate the infrared disc evolution, unless such material is itself evolving significantly on orbital time-scales. However, the data are only consistent with this interpretation; they do not rule it out. Long-term monitoring will be needed to settle this issue, given that minor planetary bodies surviving the evolution through the giant branch are likely to be found in orbits with periods of several years or longer \citep{Mustill2012}.

The detected variations have implications for previous work on dusty white dwarfs. Disc parameters derived from photometry obtained at different times may suffer wider uncertainties than reported, as the disc may have changed state between observations. Likewise, some detectable discs may have gone unnoticed if observed during a period where their flux dipped below the sensitivity limit. Revisiting previously observed systems that did not show an infrared excess may therefore be worthwhile.

The discovery of widespread time variation at dusty white dwarfs highlights the need for regular monitoring of these stars. \textit{WISE} does not reach sufficient S/N to permit the majority of the population to be studied in detail, and its six-month cadence is potentially longer than the time-scale of some of the changes. Observations at higher S/N and with shorter cadence are needed to reveal the nature of the detected variation, and to provide sufficient data to test analytic models. The delay in the launch of the \textit{James Webb Space Telescope} places this burden on \textit{Spitzer}, which is approaching the end of its nominal lifetime. It is imperative that a monitoring campaign begins in earnest before the capability is lost.

\section*{Acknowledgements}

The authors are grateful to the anonymous referee, whose comments helped improve the manuscript. This publication uses: data products from the \textit{Wide-field Infrared Survey Explorer}, a joint project of the University of California, Los Angeles and the Jet Propulsion Laboratory/California Institute of Technology, and NEOWISE, a project of the Jet Propulsion Laboratory/California Institute of Technology, both funded by the National Aeronautics and Space Administration; the NASA/IPAC Infrared Science Archive, which is operated by the Jet Propulsion Laboratory, California Institute of Technology, under contract with the National Aeronautics and Space Administration; and the SIMBAD database, operated at CDS, Strasbourg, France. AS and TGW acknowledge support from STFC studentships. JF acknowledges support from STFC grant ST/R000476/1.




\bibliographystyle{mnras}
\bibliography{InfraredVariability} 




%
%


\bsp	
\label{lastpage}
\end{document}